# Les interprétations de la mécanique quantique : une vue d'ensemble introductive

Thomas Boyer-Kassem

TiLPS, Université de Tilburg, Pays-Bas

**Résumé :** La mécanique quantique est une théorie physique contemporaine réputée pour ses défis au sens commun et ses paradoxes. Depuis bientôt un siècle, plusieurs interprétations de la théorie ont été proposées par les physiciens et les philosophes, offrant des images quantiques du monde, ou des métaphysiques, radicalement différentes. L'existence d'un hasard fondamental, ou d'une multitude de mondes en-dehors du nôtre, dépend ainsi de l'interprétation adoptée. Cet article, en s'appuyant sur le livre Boyer-Kassem (2015), *Qu'est-ce que la mécanique quantique ?*, présente trois principales interprétations quantiques, empiriquement équivalentes : l'interprétation dite orthodoxe, l'interprétation de Bohm, et l'interprétation des mondes multiples.

## 1. Introduction

À quoi ressemble le monde de l'infiniment petit ? Quelles sont les entités qui le peuplent et les lois qui en règlent le cours ? Existe-t-il un hasard fondamental, ou le monde est-il déterministe dans ses moindres recoins ? Il existe une théorie physique contemporaine qui permet de répondre à ces questions : la mécanique quantique[1]. Ou plutôt, elle autorise à chaque fois plusieurs réponses, car il est possible d'avoir des interprétations différentes de cette théorie, et il n'existe pas véritablement de consensus actuellement parmi les physiciens ou parmi les philosophes concernant *la* bonne interprétation quantique. Les interprétations quantiques offrent des images différentes du monde dans lequel la théorie est vraie, avec des types d'entités et de propriétés différents. Autrement dit, il n'y a pas d'accord parmi les physiciens ou les philosophes sur ce qui compose le monde de l'infiniment petit ! Toutefois, ces différentes interprétations sont empiriquement équivalentes au sens où elles ne peuvent être départagées par l'expérience. Cela signifie que les physiciens ne sont pas en désaccord sur les prédictions expérimentales – les interprétations de la mécanique quantique ne sont pas des théories concurrentes en un sens fort. Le fait qu'elles soient toutes autant adéquates empiriquement explique en partie l'absence de consensus parmi les spécialistes.

Les interprétations proposent des images quantiques du monde radicalement différentes. En quoi cette pluralité d'interprétations et d'images du monde est-elle un problème philosophique ? Elle l'est pour tout projet métaphysique, qui s'attache à dire quels sont les objets, les catégories, les propriétés de notre monde. Par exemple : existe-t-il plusieurs mondes parallèles ? La Nature est-elle régie par du hasard ? Les interprétations quantiques peuvent fournir des explications différentes d'un même phénomène : devrait-on renoncer à l'idée qu'une explication puisse être *la meilleure* ?

Le but de cet article est d'introduire aux principales interprétations quantiques qui concentrent l'essentiel des discussions philosophiques, et plus particulièrement de préciser l'image du monde quantique que chacune offre. Trois interprétations, parmi les plus populaires aujourd'hui chez les physiciens et les philosophes de la physique, sont considérées ici : l'interprétation dite « orthodoxe », l'interprétation de Bohm, et l'interprétation des mondes multiples. J'emprunte dans cet article de généreux extraits à mon ouvrage *Qu'est-ce que la mécanique quantique ?*, paru chez

---

1 Cet article se limite à la mécanique quantique non-relativiste, c'est-à-dire dans laquelle les effets de la relativité ne sont pas pris en compte. La théorie qui les prend en compte est la théorie quantique des champs.

Vrin en 2015, auquel le lecteur est renvoyé pour une présentation plus détaillée des interprétations quantiques, mais aussi de la non-localité et du théorème de Bell.

**2. L'interprétation orthodoxe**

Considérons tout d'abord l'interprétation que l'on trouve, au moins implicitement, dans la très grande majorité des manuels contemporains de mécanique quantique[2], et qui est enseignée presque partout dans le monde universitaire. Pour cette raison, on l'appelle généralement l'interprétation « orthodoxe ». Elle s'est imposée dès les années 1930, notamment à la suite des travaux de Bohr et de Heisenberg.

*Formulation de la théorie*

La mécanique quantique requiert que soit tout d'abord précisé le système physique considéré, par exemple un atome[3]. La théorie attribue à ce système un certain état mathématique, appelé aussi fonction d'onde, traditionnellement noté entre les symboles « | » et « > ». Un état qui décrit un atome qui se trouve à un certain endroit, ici, sera par exemple noté « | ici > ». L'état quantique permet de faire des prédictions expérimentales. En mécanique quantique, les prédictions ont la particularité d'être probabilistes : la théorie donne seulement la chance que tel ou tel résultat soit obtenu. À la question : « quelle sera la position de l'atome à tel moment ? », la mécanique quantique peut par exemple prédire qu'il y a 1 chance sur 2 qu'il se trouve ici et 1 chance sur 2 qu'il se trouve là, comme c'est le cas avec l'état « | ici > + | là > ». On parle alors d'état superposé entre l'état « | ici > » et l'état « | là > ».

Si aucune mesure n'est effectuée sur le système, son état évolue sans à-coup particulier, selon une équation dite « de Schrödinger ». Si une mesure est effectuée, l'état du système peut changer brusquement lors de cette mesure. En fonction du résultat obtenu lors de la mesure, un nouvel état est attribué au système. Dans le cas le plus simple, il s'agit de l'état correspondant au résultat de la mesure, et un système qui était dans l'état « | ici > + | là > » qui est mesuré ici verra son état projeté sur « | ici > ». L'interprétation orthodoxe considère donc, de façon générale, qu'une mesure ne *révèle* pas l'état du système, mais le modifie, et ce de façon aléatoire.

*L'image orthodoxe du monde*

Précisons maintenant l'image du monde selon l'interprétation orthodoxe de la mécanique quantique. Tout d'abord, l'état du système (ou la fonction d'onde) n'est pas considéré comme une entité du monde, ou comme référant ou correspondant à un objet du monde. Il est seulement considéré comme un outil prédictif, qui permet de calculer les différentes probabilités de mesure. Ce ne sont pas les états des systèmes, mais les systèmes quantiques eux-mêmes qui ont le statut d'entités, au sens où ils composent l'image du monde et peuvent recevoir des propriétés. Le monde orthodoxe se compose d'électrons, de photons ou de molécules.

Une autre caractéristique du monde orthodoxe est certainement moins intuitive : un système n'a pas toujours de propriété. Dans de nombreux cas, l'interprétation orthodoxe n'attribue pas de position, de vitesse ou d'énergie à un atome, ou elle dit que ces propriétés ne sont pas définies. Plus précisément, un système est considéré comme ayant une propriété seulement lorsque le résultat de mesure peut être prédit avec certitude. C'est par exemple le cas immédiatement après une mesure. En effet, la mesure a réduit l'état du système sur un état propre, correspondant au résultat de la mesure. Par exemple, l'état | ici > + | là > a été réduit sur l'état | là >, si le système a été mesuré là. À

---

[2] En français, on consultera par exemple C. Cohen-Tannoudji *et al.* 1973.
[3] Les atomes sont des constituants de la matière, de taille environ un million de fois plus petit qu'un millimètre. On trouve des atomes d'oxygène, de carbone ou d'azote par exemple. L'origine grecque du mot « atome » signifie qu'il ne peut être divisé, mais les physiciens se sont ensuite rendus ce n'est pas exact : un atome se compose d'un noyau et d'électrons, qui peuvent être séparés.

un tel état propre, la mécanique quantique associe une propriété, dans notre exemple la position, avec la valeur « là ». Cela est cohérent et d'une certaine façon bienvenu : juste après une mesure où le système a été trouvé là, on peut encore dire qu'il est là. A contrario, on dit que le système avec l'état | ici > + | là > n'a pas de position, parce que la prédiction quantique n'est pas certaine.

Comment doit-on comprendre les probabilités qui sont au cœur des prédictions de la mécanique quantique ? Selon l'interprétation orthodoxe, ces probabilités sont le signe d'un hasard fondamental ou, pour le dire autrement, le monde est *indéterministe*. Le hasard survient lors d'une mesure, au moment de la réduction que subit l'état du système. Cette réduction est aléatoire : rien, au sein du système quantique lui-même ou de l'appareil de mesure, ne pré-détermine le résultat de la mesure et la projection de l'état suivant tel ou tel nouvel état. Ce qui est fixé, en revanche, c'est la régularité statistique avec laquelle les différents résultats sont obtenus, pour un état donné. Par exemple, pour un système dans l'état | ici > + | là >, on l'observe expérimentalement effectivement en moyenne 1 fois sur 2 ici et 1 fois sur 2 là[4].

Comme le résultat de la mesure n'est déterminé par rien de plus que cette régularité statistique, on dit que les probabilités employées dans les prédictions de la théorie sont à interpréter objectivement, c'est-à-dire qu'elles représentent un hasard objectif, réel. Dieu joue vraiment aux dés, pour ainsi dire. Même lui ne peut dire, avant le résultat de mesure, si le système va effectivement être trouvé ici ou là. Les probabilités quantiques ne reflètent donc pas une ignorance de notre part, et l'état quantique décrit complètement le système. C'est en ce sens que les probabilités quantiques représentent, selon l'interprétation orthodoxe, un hasard fondamental et inhérent à notre monde. Ce hasard se traduit par la perturbation fondamentale et incontrôlable qui provient de la mesure (ou de l'appareil de mesure) sur le système quantique.

Il est important de noter que ce caractère indéterministe ne concerne qu'une seule partie de la dynamique des systèmes quantiques : la réduction de l'état lors d'une mesure. L'équation de Schrödinger qui régit l'évolution temporelle de l'état, hors mesure, est quant à elle tout à fait déterministe. Il n'y a aucun hasard dans l'évolution de l'état *entre* deux mesures.

L'existence de deux règles d'évolution distinctes (réduction de l'état, équation de Schrödinger) suppose la distinction entre les interactions qui sont à considérer comme des mesures et celles qui n'en sont pas. Cela suppose par conséquent de distinguer d'une part ce qui joue le rôle d'un appareil de mesure, responsable des premières, et d'autre part tout le reste du monde, traité quantiquement, responsable des secondes. Cette séparation entre un appareil de mesure classique et un monde quantique est au cœur de la mécanique quantique orthodoxe, qui ne peut traiter tout le monde quantiquement : une partie du monde doit être classique pour pouvoir interagir avec le système quantique et être à même d'enregistrer un résultat de mesure. Même si cette séparation peut changer en fonction de l'expérience[5], son existence est indispensable pour l'interprétation orthodoxe de la mécanique quantique. L'image orthodoxe du monde est toujours divisée en deux, l'une classique, l'autre quantique.

*Le problème de la mesure*

L'interprétation orthodoxe est largement acceptée dans la communauté scientifique en dépit d'un problème conceptuel, appelé traditionnellement « problème de la mesure », qui ronge cette interprétation depuis ses débuts, sous différentes versions[6]. Par problème conceptuel, il faut entendre l'existence d'un problème de cohérence interne concernant la formulation de la théorie et son interprétation. Cependant, ce problème n'empêche absolument pas la théorie d'être utilisée et appliquée avec succès par les physiciens. Selon une formule célèbre de Bell, à propos de la mécanique quantique orthodoxe : « à toutes fins pratiques, tout va bien[7] ». C'est d'ailleurs pour cette raison que le problème de la mesure est souvent ignoré par des physiciens ayant une approche pragmatique. Il n'en reste pas moins qu'un problème existe concernant la formulation précise de la théorie.

---

4   De tels états quantiques sont différents d'états classiques d'ignorance où le physicien attribue une probabilité de 50 % pour ici et 50 % pour là, car ils permettent par exemple de donner lieu à des interférences. Cf. par exemple Boyer-Kassem (2015, chap. 3).

5   La limite entre les parties classique et quantique du monde n'est pas définitive ; par exemple, ce qui était considéré comme un appareil de mesure peut être ensuite traité quantiquement par le physicien, dès lors qu'une autre partie du monde est considérée classiquement, et joue le rôle d'un autre appareil de mesure.

6   Les références classiques sur ce sujet incluent Albert (1992, chap. 4), Bell (1990), Krips (2013), Wallace (2008).

Le problème de la mesure naît de l'existence de deux règles d'évolution pour l'état du système, l'équation de Schrödinger et la réduction de l'état. Ces lois sont incompatibles et ne peuvent s'appliquer simultanément : la première est déterministe et continue, la seconde est indéterministe et discontinue. Le problème est que la théorie ne définit pas les circonstances dans lesquelles les deux règles différentes s'appliquent. Autrement dit, le terme de « mesure », qui est au cœur des axiomes de la théorie, n'est pas défini. La mécanique quantique orthodoxe ne donne pas de limite à ce qui vaut comme mesure. Elle est, selon les termes de Bell, « ambiguë par principe[8] ». Cette frontière peut changer au gré des utilisations de la théorie, lui donnant un regrettable « caractère fuyant[9] ».

Certaines tentatives de résolution du problème ont été proposées, mais elles n'améliorent pas le flou initial : il en va ainsi des prescriptions selon lesquelles l'appareil de mesure doit être « macroscopique », présenter un comportement « irréversible », être lié à un « observateur », etc. Ces concepts ne sont pas particulièrement mieux définis que celui de « mesure » qui figure dans la formulation orthodoxe de la théorie.

Répétons-le : le problème est d'ordre conceptuel et non pas d'ordre empirique. Les physiciens n'ont aucune difficulté à se servir de la théorie pour en tirer des prédictions, et ils savent d'expérience comment délimiter l'appareil de mesure et le système quantique afin d'obtenir la précision requise. La mécanique quantique est parfaitement convenable d'un point de vue pragmatique. Le problème est seulement d'énoncer la théorie clairement, de façon cohérente et sans ambiguïté.

Ce problème a été appelé « problème de la mesure » à cause de la formulation qu'il a prise initialement dans le cadre de l'interprétation orthodoxe : il porte sur la définition de ce qu'est une mesure. De façon plus générale, le problème de la mesure consiste à proposer une interprétation satisfaisante de la mécanique quantique (et, éventuellement, une nouvelle formulation de la théorie), qui soit en accord avec les résultats empiriques. Puisque l'interprétation orthodoxe souffre d'un problème conceptuel, il apparaît légitime d'avancer d'autres interprétations de la théorie. Aussi le problème de la mesure est-il généralement tenu pour l'origine de la diversité des interprétations quantiques.

## 3. L'interprétation de Bohm

Dans la mécanique quantique orthodoxe, les prédictions probabilistes sont interprétées comme reflétant un indéterminisme fondamental, et on considère que l'état quantique fournit une description complète du système. Une telle interprétation a longtemps rencontré des résistances. N'est-il pas possible de dépasser le caractère probabiliste des prédictions, et d'être capable de prédire assurément le résultat d'une mesure ? Dans ce but, ne peut-on pas compléter l'état de la mécanique quantique orthodoxe par d'autres variables « cachées », qui détermineraient ce résultat ? Alors, les probabilités quantiques seraient seulement le reflet de notre ignorance vis-à-vis du détail de ces variables additionnelles.

*Formulation de la théorie*

L'interprétation de Bohm[10] peut être considérée comme le résultat d'une tentative de compléter la mécanique quantique orthodoxe. En plus de la fonction d'onde, elle décrit un système quantique avec des « variables cachées », en l'occurrence les positions des particules. Ces dernières ont toujours une valeur précise à chaque instant et elles déterminent le résultat d'une mesure. La mécanique bohmienne est ainsi déterministe et les probabilités des prédictions théoriques ne sont que le reflet d'une ignorance de notre part vis-à-vis de ces variables cachées. Néanmoins, l'arrangement théorique de ces variables cachées est tel que les prédictions de la mécanique bohmienne sont exactement les mêmes que celles de la mécanique quantique orthodoxe. Ainsi, compléter la théorie avec certaines variables en décrivant une histoire en-dessous du formalisme orthodoxe, et parvenir à améliorer les prédictions empiriques, sont deux choses distinctes ; la mécanique bohmienne fait la première, mais pas la seconde. L'interprétation de Bohm suppose que l'état quantique, ou la fonction d'onde, évolue toujours selon l'équation de Schrödinger ; autrement dit, il n'existe pas de postulat de réduction[11].

---

7   « [IT] IS JUST FINE FOR ALL PRACTICAL PURPOSES », J. S. Bell (1990, p. 33).
8   Bell (1990, p. 35).
9   Bell (1987, p. 188).
10  Sur l'interprétation de Bohm, voir notamment Albert (1992, chap. 7), Bohm (1952), Dürr et Teufel (2009), Goldstein (2009), Wallace (2008, sec. 6).

*L'image bohmienne du monde*

Précisons en quoi consiste l'image du monde selon l'interprétation bohmienne, et tout d'abord ce que sont les entités qu'elle considère. Il en existe deux types : la fonction d'onde d'une part, et les particules d'autre part.

La fonction d'onde, tout d'abord, est considérée dans sa dimension spatiale seulement, c'est-à-dire comme une fonction qui associe à chaque point de l'espace un nombre, à un instant donné (un peu comme on peut attribuer à chaque point de l'espace une température). Cette fonction d'onde est considérée comme une entité authentiquement physique ; ces nombres en chaque point de l'espace renvoient à quelque chose de réel et d'objectif, qui existe bel et bien. La fonction d'onde bohmienne n'a donc rien à voir avec la simple représentation mathématique, utile dans les calculs, de la mécanique quantique orthodoxe.

Les particules constituent la seconde sorte d'entités que l'interprétation bohmienne considère. Ainsi que l'énonce un manuel de mécanique bohmienne, « chaque fois que vous dites 'particule', pensez-le vraiment ! »[12]. Cela signifie notamment qu'il faut prendre le terme en un sens traditionnel et classique, comme référant à un objet qui a une position précise à chaque instant.

Ces particules sont fondamentales en mécanique quantique bohmienne dans la mesure où toutes les autres grandeurs mesurables – vitesse ou impulsion, énergie... – peuvent s'exprimer au moyen de la position des particules. En effet, les bohmiens insistent sur le fait que toute mesure se ramène toujours *in fine* à la détermination de positions : position d'une aiguille d'un instrument, position d'un atome en sortie d'un appareil de mesure, position d'un photon sur notre rétine, etc.

L'image bohmienne du monde est déterministe. D'une part, la fonction d'onde évolue selon l'équation de Schrödinger, dont on a dit qu'elle est déterministe ; aucun hasard n'entre en compte, et la fonction d'onde ne subit jamais de projection aléatoire. D'autre part, la position des particules est donnée par une équation qui fait intervenir seulement la fonction d'onde, sans aucune notion de hasard non plus. Pour un électron décrit par l'état | ici > + | là >, l'interprétation bohmienne affirme que la position de l'électron a une valeur déterminée (rappelons que l'interprétation orthodoxe refuse de dire que l'électron a une position dans cet état). Le fait que l'électron soit véritablement mesuré ici ou là n'est pas dû à un hasard fondamental.

En revanche, le monde bohmien nous *apparaît* indéterministe, car nous n'avons pas accès aux positions des particules, comme par exemple celle de l'électron lorsqu'il est dans une superposition spatiale. Sans connaissance de la valeur de ces positions, nous ne pouvons dire ni où se trouvent exactement les particules, ni où elles se trouveront à un instant ultérieur. Cependant, nous ne sommes pas complètement démunis. La fonction d'onde, tout d'abord, peut être connue précisément. Par ailleurs, la théorie permet d'affirmer (à partir du postulat de l'équilibre quantique) que la densité de particules dans l'espace dépend directement de la fonction d'onde. Autrement dit, si la fonction d'onde est nulle ici, alors il ne peut pas y avoir de particules, et si elle a une grande valeur là, alors il y aura plus de chance d'y trouver des particules.

---

11  Après une mesure, l'état bohmien aura-t-il alors une valeur différente par rapport à l'état orthodoxe ? On montre que, à toutes fins pratiques et calculatoires, on peut considérer que la fonction d'onde bohmienne évolue en fait de la même façon que celle de la mécanique quantique orthodoxe. Cela est à l'origine de l'équivalence empirique entre des deux interprétations.
12  Dürr et Teufel (2009, p. v et 7).

Aussi, les probabilités de la mécanique quantique prennent avec l'interprétation bohmienne un tout autre sens qu'avec l'interprétation orthodoxe. Les probabilités reflètent seulement une ignorance de notre part vis-à-vis d'une histoire sous-jacente qui détermine le cours des événements. Ne connaissant que la densité moyenne des particules, nous sommes réduits à fournir des prédictions moyennes. Comme les probabilités reflètent ici non pas un hasard objectif, mais une méconnaissance de notre part, on dit qu'elles sont à interpréter de façon épistémique.

**4. L'interprétation des mondes multiples**

Une autre interprétation de la mécanique quantique a les faveurs de nombreux physiciens et philosophes des sciences. Il s'agit de l'interprétation proposée par Everett en 1957 et qui est aussi appelée l'interprétation des mondes multiples (cette dénomination est prise ici pour synonyme d' « interprétation d'Everett »)[13].

De la mécanique quantique orthodoxe, l'interprétation everettienne supprime le postulat de projection de la fonction d'onde lors d'une mesure, et ne garde que l'équation de Schrödinger. Cette dernière est la seule et vraie équation du mouvement, à laquelle obéit tout état quantique. Il n'y a plus d'ambiguïté dans l'application des lois quantiques, ni dans la définition de ce qu'est une « mesure ».

L'interprétation d'Everett considère qu'il existe une seule entité fondamentale, l'état ou la fonction d'onde (de tout l'univers). Cet objet mathématique est interprété comme une entité physique putative. C'est l'univers lui-même, en tant qu'il est une fonction d'onde, qui évolue selon l'équation de Schrödinger.

Une nouveauté radicale est introduite par l'interprétation d'Everett : certains états quantiques s'interprètent à l'aide de plusieurs mondes, au sein de cet univers[14]. Considérons un atome décrit par l'état | ici > + | là >, dont on mesure la position. L'interprétation orthodoxe dit que les deux résultats possibles que l'on peut obtenir sont « ici » ou « là », et que l'on en obtient un seul. L'interprétation des mondes multiples, de son côté, affirme que les deux résultats sont obtenus, chacun dans un monde. Comme ce résultat dépend du monde auquel on se restreint, cela conduit à définir les états ou les faits *relativement* à un observateur – d'où le nom de « formulation de l'état relatif » initialement donné par Everett. Si on parle parfois *du* résultat d'une mesure, c'est en fait par abus de langage, en omettant de préciser que cela se comprend relativement à *un* monde particulier. Pour l'univers dans son ensemble, il n'existe pas de fait à propos du résultat de mesure de la position de l'atome ; pour l'univers entier, l'atome n'a pas été mesuré « ici » ou « là ».

Les interactions quantiques ayant lieu entre les moindres électrons ou atomes suscitent un processus d'embranchement qui multiplie à chaque instant le nombre de mondes, de sorte qu'il existe un nombre extraordinairement grand de mondes. Ce qui existe pour un everettien, c'est donc une myriade de mondes. Dans chacun de ces mondes, les grandeurs ont toujours des valeurs ; ces mondes sont donc d'apparence classique, et ils se composent d'objets (macroscopiques) qui sont dans des états définis. Les différents mondes évoluent indépendamment les uns des autres. En particulier, les autres mondes sont inobservables depuis un monde particulier, ce qui explique pourquoi nous avons toujours l'impression qu'il n'existe qu'un seul monde.

L'univers everettien est déterministe. En effet, la fonction d'onde de l'univers obéit à l'équation de Schrödinger, dont on a dit qu'elle est une équation déterministe. L'avenir n'est pas incertain, puisque tous les résultats de mesures possibles se produiront toujours. En revanche, les individus dans les différents mondes ont des expériences psychologiques différentes. Le cours du monde leur apparaît indéterministe, dans la mesure où ils n'ont accès qu'à un seul monde. Pour l'interprétation des mondes multiples, les probabilités associées aux résultats correspondent aux paris que peuvent faire les individus. Comme elles n'expriment pas une connaissance incomplète de leur part, elles ne sont pas subjectives, mais objectives[15].

---

13 Concernant cette interprétation, voir notamment sur Albert (1992, chap. 6), Barrett (2011), Everett (1957), Vaidman (2008), Wallace (2008, sec. 4).
14 Les détails mathématiques de ces états ne sont pas présentés ici, en raison de contraintes d'espace. Nous renvoyons à Boyer-Kassem (2015, chap. 5).

Noter que l'interprétation d'Everett permet à la mécanique quantique de s'appliquer à l'ensemble de l'univers. Contrairement à l'interprétation orthodoxe, elle ne suppose pas de division entre un « système », distingué d'un « observateur » qui constate les résultats de mesures. L'univers everettien n'est pas séparé entre une partie classique et une partie quantique.

**5. Conclusion**

La mécanique quantique est une théorie physique qui admet plusieurs interprétations, lesquelles dessinent des images du monde radicalement différentes, mais ne peuvent être distinguées empiriquement. Cela signifie qu'aucune expérience réalisable ne permettra jamais de trancher entre, par exemple, l'idée d'un monde déterministe à la Bohm, dans lequel aucun hasard n'intervient dans le cours des événements, ou l'idée d'un monde indéterministe, comme le veut l'interprétation orthodoxe, au sein duquel un hasard fondamental joue un rôle presque à chaque instant. Dès lors que l'image du monde quantique est prise au sérieux, l'expérience ne permet pas de trancher la question de savoir si le hasard pur existe ou non dans notre monde.

---

15  Elles expriment des contraintes auxquelles sont soumises tous les agents rationnels. *Cf.* par exemple Wallace (2008, sec. 4.6).

**Bibliographie**

David Z. Albert, *Quantum Mechanics and Experience*, Cambridge (MA) et London, Harvard University Press, 1992.

Jeffrey Barrett, « Everett's Relative-State Formulation of Quantum Mechanics », dans dans E. N. Zalta (éd), *The Stanford Encyclopedia of Philosophy*, http://plato.stanford.edu/archives/spr2011/entries/qm-everett/, 2011.

John S. Bell, *Speakable and Unspeakable in Quantum Mechanics*, Cambridge, Cambridge University Press, 1987.

John S. Bell, « Against 'measurement' », *Physics World*, août 1990, p. 33-40.

David Bohm, « A Suggested Interpretation of Quantum Theory in terms of 'Hidden' Variables », *Physical Review*, 85, 1952, p. 166-193.

Thomas Boyer-Kassem, *Qu'est-ce que la mécanique quantique ?*, Paris, Vrin, coll. « Chemins Philosophiques », 2015.

Claude Cohen-Tannoudji, Bernard Diu et Franck Laloë, *Mécanique Quantique*, Tome 1, Paris, Hermann, 1973/1998.

Detlef Dürr et Stefan Teufel, *Bohmian Mechanics: The Physics and Mathematics of Quantum Theory*, Berlin et Heidelberg, Springer-Verlag, 2009.

Sheldon Goldstein, « Bohmian Mechanics », dans E. N. Zalta (éd), *The Stanford Encyclopedia of Philosophy*, http://plato.stanford.edu/archives/spr2009/entries/qm-bohm/, 2009

Hugh Everett, « 'Relative State' Formulation of Quantum Mechanics », *Reviews of Modern Physics*, 29, 1957, p. 454-462.

Henry Krips, « Measurement in Quantum Theory », dans E. N. Zalta (éd), *The Stanford Encyclopedia of Philosophy*, http://plato.stanford.edu/archives/fall2013/entries/qt-measurement/, 2013.

Lev Vaidman, « Many-Worlds Interpretation of Quantum Mechanics », dans dans E. N. Zalta (éd), *The Stanford Encyclopedia of Philosophy*, http://plato.stanford.edu/archives/fall2008/entries/qm-manyworlds/, 2008.

David Wallace, « The Quantum Measurement Problem: State of Play », dans D. Rickles (éd.), *The Ashgate Companion to Contemporary Philosophy of Physics*, Aldershot, Ashgate Publishing, 2008, p. 16-98, disponible en prépublication à http://arxiv.org/abs/0712.0149.